\documentclass{article}

\usepackage[english]{babel}

\usepackage{amsmath}
\usepackage{graphicx}
\usepackage[colorlinks=true, allcolors=blue]{hyperref}
\usepackage{amsfonts}
\usepackage{listings}
\usepackage[T1]{fontenc}
\usepackage{caption}
\usepackage{authblk}   
\usepackage{url}      
\usepackage{fancyhdr}  
\usepackage{float}
\usepackage{cite}    
\usepackage{url}    
\usepackage[margin=1in]{geometry}

\title{Applications Of Zero-Knowledge Proofs On Bitcoin
}

\author{
    {\LARGE Yusuf Ozmiş} \\
    Yildiz Technical University \\
    \texttt{yusuf.ozmis@std.yildiz.edu.tr}
}

\date{}

\begin{document}
\maketitle

\begin{abstract}
This paper explores how zero-knowledge proofs can enhance Bitcoin's functionality and privacy. First, we consider Proof-of-Reserve schemes: by using zk-STARKs, a custodian can prove its Bitcoin holdings are more than a predefined threshold X, without revealing addresses or actual balances. We outline a STARK-based protocol for Bitcoin UTXOs and discuss its efficiency. Second, we examine ZK Light Clients, where a mobile or lightweight device verifies Bitcoin's proof-of-work chain using succinct proofs. We propose a protocol for generating and verifying a STARK-based proof of a chain of block headers, enabling trust-minimized client operation. Third, we explore Privacy-Preserving Rollups via BitVM: leveraging BitVM, we design a conceptual rollup that keeps transaction data confidential using zero-knowledge proofs. In each case, we analyze security, compare with existing approaches, and discuss implementation considerations. Our contributions include the design of concrete protocols adapted to Bitcoin's UTXO model and an assessment of their practicality. The results suggest that while ZK proofs can bring powerful features (e.g., on-chain reserve audits, trustless light clients, and private layer-2 execution) to Bitcoin, each application requires careful trade-offs in efficiency and trust assumptions.
\end{abstract}

\section{Introduction}

Zero-knowledge proofs gained significant attention in the blockchain ecosystem for their ability to provide privacy and scalability without sacrificing verifiability. Early use cases of zero-knowledge proofs were focused on privacy-preserving transactions, such as Zero-Coin\cite{miers2013zerocoin}, which later became Zcash\cite{bensasson2014zerocash}.  On Ethereum\cite{wood2014ethereum}, ZKPs have evolved to support advanced use cases including anonymous identity systems, Private Voting\cite{majeed2018evoting}, and verifiable computation, which evolved into zero-knowledge rollups such as Starknet\cite{starkware2025starknet} and Scroll\cite{scroll2025zkevm}. These applications benefit from Ethereum's smart contract capabilities, enabling expressive on-chain verification logic.

In contrast, Bitcoin’s\cite{nakamoto2008bitcoin} design favors simplicity over programmability. It lacks a Turing-complete scripting language and opcodes necessary to verify zero-knowledge proofs on-chain. This resulted in very limited integration of zero-knowledge proofs on Bitcoin. Hence, privacy on Bitcoin has largely relied on non-cryptographic methods such as CoinJoin\cite{coinjoin2013wiki}, which masks transaction links through coordinated mixing, and protocols like PayJoin\cite{bitcoinwiki2025payjoin} and JoinMarket\cite{btcpay2025joinmarket}. While effective against basic heuristics, these methods do not offer strong cryptographic guarantees of privacy and remain susceptible to de-anonymization through statistical or timing analysis.

We demonstrate how zero-knowledge proofs can be utilized on Bitcoin. Specifically, we introduce a proof-of-reserve scheme that utilizes zero-knowledge proofs to enable users to prove the ownership of a given UTXO exceeding a threshold value, without revealing transaction details. Unlike centralized or Merkle-based approaches, our system achieves soundness and privacy using succinct cryptographic proofs. Additionally, we explore a zero-knowledge consensus proof that allows Bitcoin light clients to verify Bitcoin’s consensus without re-executing. We also explore privacy-preserving use cases, specifically a Privacy-rollup via BitVM\cite{linus2023bitvm}.

While Zero-knowledge proofs have been widely adopted in other blockchain ecosystems, their applications on Bitcoin remain limited by Bitcoin's conservative design principles. Our work bridges this gap by showing applications of zero-knowledge proofs on Bitcoin, such as proof-of-reserve, consensus proof, and privacy-preserving use cases. While some of these applications can already be built, such as proof of reserves, others require extending Bitcoin’s opcode set.

\section{Preliminaries}

This section provides an introduction to concepts that will be frequently referenced throughout the paper. 

\subsection{Rollups}

The scalability of a blockchain refers to its ability to maintain a certain level of performance and security while handling a growing volume of transactions.  Various methods such as Sharding\cite{kommineni2022sharding} and novel consensus protocols\cite{yin2019metastability} have been proposed to enhance scalability, but often compromise on decentralization. To address these limitations, Layer 2 solutions emerged with the goal of reducing the execution load on the Layer 1, primary blockchain, without sacrificing safety or decentralization. However, Layer 2 solutions came with their own trade-offs: payment channels\cite{papadis2020pcn} lack programmability, while sidechains and Plasma\cite{poon2017plasma} introduce additional trust assumptions. 

One of the most promising scaling solutions is Rollups, introduced to eliminate the trade-offs that other Layer 2’s have. Rollups are blockchains that utilize the Layer 1 blockchain as a Data Availability Layer. Using Layer 1 as a Data Availability Layer ensures that Layer 1 full nodes can reconstruct the rollups' state. Rollups may also choose to inherit Layer 1 settlement guarantees, which means verifying rollups’ state changes on Layer 1. Settlement can be optimistic or via zero-knowledge proofs. 
Rollups are categorized based on their choices for the Data Availability and Settlement Layer. For example, a rollup that uses Bitcoin as its Data Availability Layer is referred to as a Bitcoin rollup \cite{lightclient2022rollups}. The settlement approach further defines the rollup, categorizing it as optimistic, zero-knowledge, or sovereign based on how it ensures the validity and finality of state changes.

\subsection{BitVM}

BitVM is a computing paradigm to express Turing Complete contracts on Bitcoin without changing its consensus rules. BitVM allows anyone to challenge the execution of a contract if convinced that computation is invalid, making it an optimistic scheme. 

Even though BitVM can be used to verify any Turing-complete contract, most use cases are not viable due to Bitcoin transaction costs, as disproving the execution of a contract will require a set of transactions. One of the most viable use cases for BitVM is to verify zero-knowledge proofs optimistically, making Zero-Knowledge Rollups on Bitcoin possible. 

\subsection{Zero-Knowledge Proofs}

Zero-knowledge proofs (ZKPs) are cryptographic methods that allow one to prove the validity of a statement to another(verifier), without revealing any additional information. ZKPs enable a wide range of applications, including verifiable computation and private payments. 

Zero-knowledge proofs are defined by three core properties which are completeness, soundness, and zero-knowledge:

\begin{itemize}
    \item \textbf{Completeness:} if a statement is true, an honest prover can convince an honest verifier. 
    \item \textbf{Soundness:} if a statement is false, a dishonest prover cannot convince an honest verifier. 
    \item \textbf{Zero-Knowledge:} no information other than the validity of the statement is revealed.
\end{itemize}

The completeness property ensures that if a statement is true, an honest prover can convince an honest verifier. The soundness property guarantees that if a statement is false, a dishonest prover cannot convince an honest verifier.  The zero-knowledge property ensures that no information other than the validity of the statement is revealed. 

\section{Proof-of-Reserves Protocol}

We consider a scenario where a prover (e.g. an exchange or user) wants to prove the ownership of a Bitcoin UTXO with an amount greater than X, without revealing his address, and the exact amount. Such proof is particularly relevant for systems involving wrapped Bitcoin or Bitcoin-backed stablecoins, where client funds are held on-chain by the custodians. A zero-knowledge proof-of-reserve enables verification that the underlying UTXO remains unspent and that the backing reserves exceed the amount of issued tokens, all without compromising custodian’s privacy.

We assume the standard Bitcoin security model: a majority of computational power is honest, so the longest chain represents the true ledger. The adversary (dishonest prover) may try to generate a proof claiming a fake UTXO or oversize value. Our protocol must ensure soundness (no false proofs) under standard cryptographic assumptions.

\begin{itemize}
    \item \textbf{Threat Model:}  We assume the verifier trusts Bitcoin’s proof-of-work consensus (i.e., trusts that the majority is honest) and has the Bitcoin block headers (e.g. runs a Bitcoin light client). The prover may be malicious and try to prove a non-existent or small UTXO as large enough, or forge knowledge of a private key. We aim for the usual ZK guarantees: a dishonest prover cannot convince the verifier except by breaking crypto, and an honest prover reveals only the allowed information.
    
    \item \textbf{High-Level Protocol Flow:} For proof-of-reserve, we assume the prover runs a Bitcoin full-node, having all headers and raw transactions. The prover collects a block header and a Merkle branch proving the existence of a transaction TX that has an output at index i. Inside TX, at index i, the UTXO has value v. The prover knows the private key sk, that can satisfy the spending condition of the corresponding UTXO. Both parties agree on a public threshold X. 
The prover then constructs a zk-STARK\cite{bensasson2018stark} circuit that assert the following statements:

\begin{itemize}
    \item \textbf{Existence:}  The prover shows that a transaction is included in a Bitcoin block whose header is known by the verifier. This can be done with an SPV Merkle proof.

    \item \textbf{Ownership:} The prover knows a private key that can spend the UTXO. If the scriptPubKey is, say, a P2PKH (Pay-to-PubKey-Hash), this means the prover knows the private key corresponding to the pubkey hash in the output. The proof must enforce satisfaction of the script’s requirements (e.g. signature validity) without revealing the key. To ensure this, the prover must also be able to prove the validity of  secp256k1 and Schnorr signatures, and SHA256 hashes.

    \item \textbf{Value Bound:} The value of UTXO v is above the threshold X. This can be expressed by showing v - X > 0 in zero-knowledge
\end{itemize}
    
\end{itemize}

The resulting proof, along with the block header, Merkle path, scriptPubKey, and threshold, is sent to the verifier. The verifier runs the STARK verifier; on success it only learns “the prover knows some valid UTXO that has not been spent in the block H, with an amount v > X”.

\subsection{Implementation Details}

We now describe our zero-knowledge proof-of-reserve protocol in detail. We assume the following public inputs to the verifier:

\begin{itemize}
    \item \textbf{Threshold X:} A threshold value, expressed in terms of satoshis.
    
    \item \textbf{Block Headers:} $\{H_1, H_2, \ldots, H_i\}$  with   being the block UTXO is created, containing the Merkle roots $\{R_1, R_2, \ldots, R_i\}$
\end{itemize}

Prover also has the following private inputs: 

\begin{itemize}
  \item \textbf{Raw transaction:} The prover’s private input includes the raw transaction in which the UTXO was created. The prover computes the transaction ID from the raw transaction data and extracts both the \texttt{scriptPubKey} and the output value based on output index.
  \item \textbf{Output index (vout):} The prover also receives the output index (referred to as \texttt{vout}) of the UTXO being proven.
  \item \textbf{Secret key:} User’s secret key, which satisfies the locking condition, \texttt{scriptPubKey}.
  
\end{itemize}

The prover’s goal is to prove: “(a) TX is included in a block  with root  (SPV proof), (b) $ \texttt{TX.outputs[}i\texttt{].value} \geq X $ , and (c) the prover knows the secret key that can satisfy the locking condition, and (d) The Prover has not spent the UTXO with the vout i in the blocks whose headers are $\{H_1, H_2, \ldots, H_i\}$ Without revealing TX, v, or sk.

We implement this via a zk-STARK circuit. Conceptually the circuit does the following checks:

\begin{enumerate}
  \item \textbf{Merkle Inclusion Check (SPV):}
  \begin{itemize}
    \item Compute the hash of $TX$ from the raw transaction.
    \item Use the Merkle branch $\pi$ to hash up to $R$ (public). Check that it matches $R$.\\
    This verifies the $TX$ is in the block.
  \end{itemize}

  \item \textbf{Output Extraction:}
  \begin{itemize}
    \item Parse $TX$ to extract the $i$-th output ($v$, \texttt{scriptPubKey}).
    \item Check that this output’s \texttt{scriptPubKey} is of a standard form (valid transaction type)\\
    and retrieve the contained public key hash or key $PK$.
  \end{itemize}

  \item \textbf{Threshold Check:}
  \begin{itemize}
    \item Check the UTXO’s amount $v$ is greater than some predefined threshold $X$.
  \end{itemize}

  \item \textbf{Key Knowledge Check:}
  \begin{itemize}
    \item Verify that the prover knows the private key $sk$ corresponding to $PK$. For a\\
    P2PKH script, this amounts to checking that $PK = \text{Hash(pubkey)}$ and proving\\
    knowledge of $sk$ such that $g^{sk} \bmod p = \text{PubKey}$. In practice we utilize\\
    Bitcoin’s elliptic curve, secp256k1, as part of the ZK statement (proving the validity\\
    of Elliptic Curve Multiplication $g^{sk} \bmod p = \text{PubKey}$, and $PK = \text{Hash(pubkey)}$ is a part of Proof-of-Reserve circuit).
  \end{itemize}
\end{enumerate}

\begin{figure}[H]
  \centering
  \includegraphics[width=0.8\textwidth]{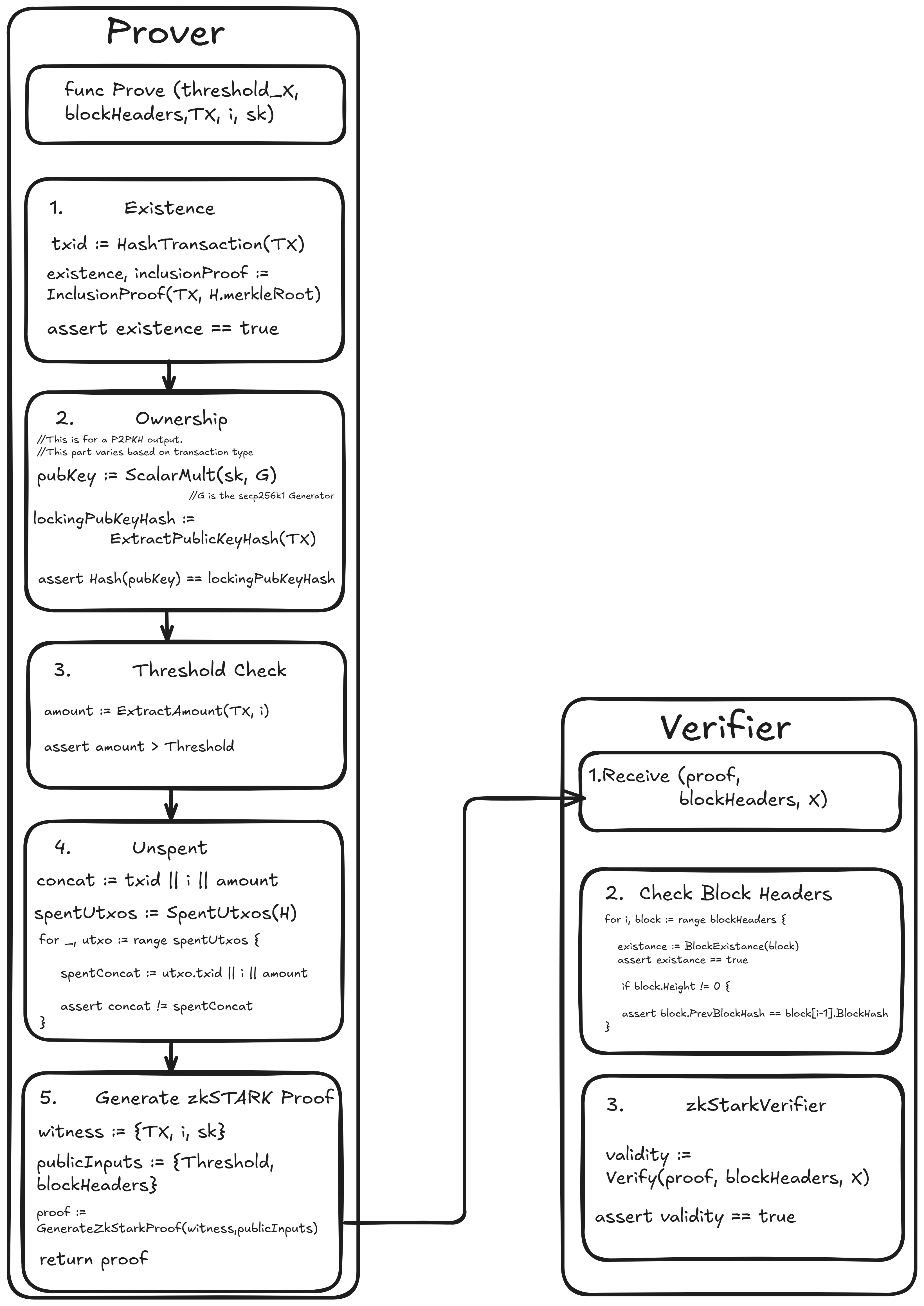}
  \caption{Overview of the zk-STARK-based Proof-of-Reserve Protocol.}
  \label{fig:yourlabel}
\end{figure}

The prover then generates a zk-STARK proof asserting that it knows such a witness (e.g., the transaction TX, secret key sk, vout i and the amount v) for which all constraints, described above, are satisfied. The verifier receives only the proof and the public inputs— namely, the block headers  $\{H_1, H_2, \ldots, H_i\}$ and threshold X, Crucially, the value v and other private data remain hidden. The soundness of the zk-STARK ensures that a valid proof could not be generated unless the constraints are satisfied ~\ref{fig:yourlabel}.

\subsection{Security Analysis of Proof-of-Reserves}

We now argue that the proof-of-reserve protocol is \textbf{complete}, \textbf{sound} and \textbf{zero-knowledge}.

\begin{itemize}
    \item \textbf{Completeness:} If the prover truly has a UTXO of value $v > X$, then an honest prover knows \textit{TX} and \textit{sk}. They can produce a correct Merkle proof and satisfy all circuit checks, so the STARK prover will generate a valid proof that the verifier will accept (by definition of the proof system).
    
    \item \textbf{Soundness:} Suppose a prover could create a convincing proof while violating the statement. There are three main types of cheating: (1) Claim a non-existent or spent UTXO, or (2) claim value $< X$ as $\> X$, or (3) fake key ownership.
    
    \begin{enumerate}
        \item \textit{Fake inclusion:} If \textit{TX} is not actually in a block with root \textit{R}, then the Merkle inclusion check would fail. A cheating prover could try to invent \textit{TX} and $\pi$, but then \textit{TXid} would have to appear in the actual Merkle tree with root \textit{R}. By collision-resistance of the hash/Merkle construction, forging this implies breaking Bitcoin’s hash security (or control of a 51\% hash power to rewrite the chain, which contradicts our consensus assumption).
        
        \item \textit{Wrong value:} If $v < X$, the prover might try to trick the circuit by mis-parsing \textit{TX} or using a malicious scriptPubKey that hides the actual value. However, the circuit explicitly extracts the integer value $v$ from the correct location in \textit{TX}. Any misplacement would be detected by the arithmetic checks. Alternatively, forging $v > X$ would require finding an \textit{TX} whose extracted value is $ >X$, so a prover cannot claim more value than exists in a genuine output.
        
        \item \textit{Key forgery:} To fake knowledge of \textit{sk}, the prover would have to break the discrete-log assumption on secp256k1. The circuit enforces $pub = G^{sk}$ inside it. A prover without the real \textit{sk} cannot produce this exponentiation
    \end{enumerate}
    
    Thus any false proof would imply either (i) breaking a hash or Merkle proof (i.e. forging chain inclusion), or (ii) solving a discrete log. Both are assumed infeasible. We conclude the protocol is sound: a cheating prover has negligible chance to convince an honest verifier if no valid UTXO $> X$ exists.
    
    \item \textbf{Zero-Knowledge:} We must argue that the verifier learns nothing beyond the claim “there exists some UTXO $> X$ known to prover, and is unspent within the blocks $\{H_1, H_2, \ldots, H_l\}$”. The STARK proof reveals no details of \textit{TX}, $v$, or \textit{sk}. By definition of zk-STARKs, the proof is \textit{non-interactive} and \textit{zero-knowledge}. The verifier cannot learn the exact value $v$, the transaction id, or the private key. It only learns that $v > X$.
\end{itemize}

In summary, under standard cryptographic assumptions, the proof-of-reserve protocol is both sound and (perfectly) zero-knowledge. The consensus-light protocol (if implemented) would similarly rely on collision-resistance of hashes and difficulty of forging chain histories.

\subsection{Limitations and Future Work}

Our protocols involve several limitations:

\begin{itemize}
    \item \textbf{Proof Size and Computation:} zk-STARK proofs are typically larger than SNARKs. In our experiments, proofs were on the order of a few megabytes for a single UTXO check. For some applications, this may be heavy. However, we expect future STARK optimizations and recursive proof-composition could reduce sizes. The prover’s computation is also non-trivial (minutes per proof), although highly parallelizable.
    
    \item \textbf{No On-Chain Verification:} Currently the proof verification is off-chain. That means a Bitcoin light client could perform it locally, but the Bitcoin blockchain cannot directly verify our proofs under existing opcodes.
    
    \item \textbf{Threshold Definition:} We prove \texttt{"$>$X"} for one UTXO. A user with multiple smaller UTXOs cannot combine them in this basic protocol. Extending to \textit{solvency proof} (sum of many UTXOs $\geq X$) is more complex (requires aggregation or set commitment techniques). Future work could explore zk-aggregations or Merkle commitments of an exchange’s entire address set.
    
    \item \textbf{Script Variants:} Our proof-of-concept circuit\footnote{\url{https://github.com/yusufozmis/utxo-ownership-proof}} can only prove for P2PKH outputs. Handling other transaction types would require custom circuit code for each.

\end{itemize}

\textbf{Future Work:} On the research side, we plan to explore recursive SNARK/STARK composition to reduce proof size, expand the transaction types, and aggregation to allow multiple UTXO’s to combine.

\section{ZK LIGHT CLIENT PROTOCOL FOR BITCOIN}

A light client is a blockchain node that does not store or validate all data (e.g. full blocks) but still wants strong security guarantees. Bitcoin’s original SPV (Simplified Payment Verification) scheme allows a client to trust the longest-chain header by checking Proof-of-Work (PoW) and verifying Merkle proofs for its own transactions. However, SPV still requires downloading all block headers (80 bytes each) and trusting they form a valid PoW chain. ZK-based light clients aim to do even less work: rather than verifying each header’s hash, the client merely checks a succinct ZK proof that the chain is valid.

In summary, a ZK light client protocol allows a prover to assert the validity of Bitcoin’s State Transition Function (PoW difficulty, block structure and transaction validity), via a ZK proof. The client (verifier) only needs to store block headers and verify that proof. This allows one to achieve full-node security with significantly lower bandwidth and storage requirements.

\subsection{Protocol Design and Assumptions}

Our proposed protocol enables a prover to convince a light client that a sequence of Bitcoin blocks (or a block header) is valid under the PoW rules. The key idea is to represent the block validation logic inside the arithmetic circuit and generate a succinct proof (e.g. with a SNARK or STARK). We outline one possible design: 

\begin{itemize}
    \item \textbf{Protocol Steps (Overview):}
    \begin{enumerate}
        \item \textbf{Longest Chain:} The prover runs a Bitcoin full-node and has the \textbf{full transactional data}. It proves the longest chain it claims is valid.
        
        \item \textbf{Statement to prove:} Formulate the statement “Block header H is a valid block of checkpoint under Bitcoin consensus”. This involves checking the validity of each transaction, checking their hashes (transaction id), checking each header and checking the PoW target constraints.
        
        \item \textbf{Proof generation (Prover):} Using a ZK prover, compute a proof $\pi$ that:
        \begin{itemize}
            \item Each Bitcoin block header H satisfies the difficulty target
            \item For each $i > 0$, \quad $H_{i-1}.hash = H_i.\text{prevHash}$
            \item Difficulty rules (e.g. no difficulty tampering) are respected.
        \end{itemize}
        
        \begin{itemize}
            \item The merkle root in the block header is computed correctly
            \item Each transaction’s txid is computed correctly.
            \item Each transaction is valid
        \end{itemize}
        
        \item \textbf{Publish proof:} Send $\pi$ to the verifier.
        
        \item \textbf{Verification (Light client):} The client checks the proof $\pi$ using the verifier key. If valid, it accepts the block headers $\{H_1, H_2, \ldots, H_k\}$ as a legitimate Bitcoin chain.
    \end{enumerate}
    
    \item \textbf{Alternate (Recursive Proofs):} Instead of proving 1 to H in one go, the prover could update proofs recursively for each new block. For example, starting from the genesis block $H_0$, the prover generates $\pi$ proving header 1, then proving header 2 on top of 1, etc. Each H can itself verify linkage.
    
    \item \textbf{Verification Key Storage:} The light client needs to store a small verification key and the latest header (and possibly a block count).
\end{itemize}

This protocol assumes that the verifier runs a Bitcoin light client, as it’ll need to check the block headers prover provided. We also assume the Bitcoin PoW difficulty function and target are known publicly. We assume at least one honest prover (or aggregator) exists to generate proofs; malicious prover(s) can be checked by anyone. The model is off-chain: proofs are generated and sent to the client, not executed within Bitcoin itself.

\subsection{Current Limitations and Discussion}

Despite its promise, a ZK light client for Bitcoin faces significant challenges:

\begin{itemize}
    \item \textbf{Heavy Computation:} Proving Bitcoin’s SHA-256 based PoW is costly. Even optimized SNARK circuits for SHA-256 require millions of constraints, and STARKs must handle wide bit operations. Thus, the prover’s computation could be substantial (minutes per block). Ongoing research is needed to speed up proving SHA-256 hash function.
    
    \item \textbf{No On-Chain Verification:} Unlike Ethereum, Bitcoin cannot natively verify SNARK proofs. Thus, the protocol works off-chain only: proofs are delivered to clients by peers. Bitcoin needs to add new opcodes, such as CAT\cite{bip420}, to verify the zero-knowledge proof on-chain.
    
\end{itemize}

In summary, a ZK light client protocol is feasible in principle and can offer near-full-node security without full validation. However, it remains in active research; as of this writing, only prototypes exist (e.g. ZeroSync’s\cite{zerosync2025} work). The Bitcoin community must weigh the benefits versus the complexity and trust assumptions.

\section{Privacy Rollups With BitVM}

Bitcoin, by design, prioritizes transparency, with every transaction publicly recorded on the blockchain. While this property enhances verifiability, it inherently limits user privacy. In contrast to smart contract platforms such as Ethereum, Bitcoin has very limited programmability, making it impossible to build privacy-preserving protocols on Bitcoin. However, recent developments such as BitVM have opened the wave for new possibilities, including privacy rollups.

Privacy rollups provide several important advantages. First, they significantly reduce the amount of data posted to the base layer, as only commitments and zero-knowledge proofs are posted to the base layer, rather than full transaction data. Second, they enhance user privacy by keeping transaction details and account states confidential—known only to the prover, typically the user, while still ensuring verifiability through zero-knowledge proofs.

\subsection{Rollup Design}

Consider a rollup that maintains an off-chain state (e.g. balances, nonces etc.) and executes transactions in private. The rollup operator, referred to as sequencer, collects user transactions (which may include encrypted outputs or one-time addresses) and updates the state accordingly. User’s do not necessarily broadcast their transaction, but may use Client-Side Proving\cite{aztec2025proofgen} methods to hide their transaction data, ensuring full privacy.

Privacy-Rollups does not have to be fully private, they may provide a hybrid architecture, such as Aztec\cite{aztec2025zkrollup}, allowing users to choose whether their transaction will be public or not.

Periodically, the operator publishes a commitment to the new state, or posts the public state if hybrid rollup, on Bitcoin, along with a zero-knowledge proof that the state transition is valid (all transactions were correct).

Key features of this privacy rollup include:

\begin{itemize}
    \item \textbf{Reduced on-chain data:}  Only cryptographic commitments (state roots) and a ZK proof are posted on-chain, not the full transaction details, which significantly reduces on-chain data.
    
    \item \textbf{Privacy:} Transaction data (senders, recipients, amounts) remain private.  Users typically share only blinded commitments or encrypted notes. The ZK proof ensures validity without revealing transaction data.
    
\end{itemize}

An example workflow is:

\begin{enumerate}
    \item \textbf{State Commit:} The rollup operator, Sequencer, posts state root, or public state if its hybrid rollup, to Bitcoin.

    \item \textbf{Proof Generation:} The operator generates a SNARK proving that the new state was computed correctly from the previous state given the batch of off-chain transactions. This circuit checks, for each transaction, that inputs are available, scripts are satisfied, and outputs sum correctly, updating the Merkle root.

    \item \textbf{On-chain Posting:} The operator posts a transaction containing the new state root alongside with the ZK proof

    \item \textbf{BitVM Verification:} If any participant detects fraud, they can use the BitVM mechanism to challenge the posted state. In BitVM, the challenger invokes an on-chain interactive dispute where the operator must run parts of the verification in Bitcoin script. If the operator cannot satisfy the proof, they lose their deposit.
\end{enumerate}

\subsection{Challenges And Limitations}

\begin{itemize}
    \item \textbf{Bitcoin Script Limitations:} As of today, verifying even simple cryptographic operations in Bitcoin Script is extremely difficult. Any BitVM-based rollup must deal with these low-level constraints when expressing the SNARK verifier logic in Bitcoin script.
    
    \item \textbf{Early Stage Tech:} BitVM is still in its experimental state and lacks mature tooling and infrastructure support.
    
\end{itemize}

Despite these challenges, the conceptual design shows a way to achieve Ethereum-like privacy rollups on Bitcoin in a trust-minimized manner. As technology matures (better script capabilities, more efficient ZK primitives, and refined BitVM protocols), such constructions could become practical.

\section{Conclusion}

This paper has presented three original contributions at the intersection of zero-knowledge proofs and Bitcoin:

\begin{itemize}
    \item \textbf{zk-STARK PoR protocol} that leverages transparent, post-quantum proofs to guarantee exchange reserves without revealing private data. We detailed the protocol steps, its implementation, and its security guarantees.
    
    \item \textbf{ZK-based light client design} for Bitcoin, enabling succinct verification of proof-of-work consensus. We specified the assumptions (running a Bitcoin Light Client proof system setup), algorithmic steps, and limitations. While fully trustless on-chain SNARK verification is not yet possible in Bitcoin, our design shows how off-chain ZK proofs can significantly reduce client work. We also highlighted open issues (computational cost, verification of ZKP on Bitcoin) for future research.
    
    \item \textbf{A conceptual BitVM rollup protocol}, outlining how Bitcoin can support a privacy-preserving layer-2 via optimistic verification. We described the protocol flow, which combines off-chain ZK proofs with Bitcoin timelocks, and addressed its security and implementation challenges. Although still experimental, BitVM represents a novel path for Bitcoin scalability and privacy. Our design particularly emphasizes hiding transaction details through ZK proofs.
\end{itemize}

\section*{Acknowledgements}

I would like to thank Koray Akpınar, Janusz Grze, and Ömer Talip Akalın for their valuable reviews and insightful feedback throughout the development of this work. Their comments significantly contributed to improving the clarity and quality of this paper.

\bibliographystyle{alpha}
\bibliography{sample}

\end{document}